# Generative AI in Software Testing: Current Trends and Future Directions

**Tanish Singla and Qusay H. Mahmoud**

Department of Electrical, Computer, and Software Engineering
Ontario Tech University, Oshawa, Ontario L1G 0C5 Canada
{tanish.singla,qusay.mahmoud}@ontariotechu.net

**Abstract:** This paper investigates current software testing systems and explores how artificial intelligence, specifically Generative AI, can be integrated to enhance these systems. It begins by examining different types of AI systems and focuses on the potential of Generative AI to transform software testing processes by improving test coverage, increasing efficiency, and reducing costs. The study provides a comprehensive overview of the current applications of AI in software testing, emphasizing its significant contributions in areas such as test case generation and validation. Through an extensive literature review, it highlights how Generative AI can streamline these processes, resulting in more robust and thorough testing outcomes. The paper also examines methods to improve the efficiency of Generative AI systems, such as prompt engineering and fine-tuning. Additionally, it explores the use of AI in specific tasks, including input generation, oracle generation, data generation, test data creation, and test case prioritization. By analyzing the current landscape and identifying both the opportunities and challenges in integrating Generative AI, this paper provides valuable insights and recommendations for practitioners and researchers. It underscores the need for ongoing advancements and targeted development efforts to overcome existing hurdles and fully leverage AI's capabilities. The findings further show that with continued innovation and careful implementation, Generative AI has the potential to significantly enhance the efficiency, effectiveness, and reliability of software testing, particularly in the rapidly evolving field of IoT testing.

**Keywords:** Software Testing Life Cycle; Natural Language Processing; Fine-tuning models; Prompt Engineering; Testing Oracles; Generative AI; Software Development Life Cycle

## 1. Introduction

Software development processes are referred to as the procedures followed when designing a software project [1], typically including key stages such as planning, analysis, design, coding, testing, and maintenance [2]. The initial stage, planning, involves considering the requirements and needs of users. Following this is the analysis phase, where timelines are estimated and potential risks are identified, often using diagrams like UML and activity diagrams. The design phase translates these plans into a structured blueprint, detailing the project's algorithm, interface, and technical specifications. The coding or implementation stage actualizes these designs into a functioning code environment. Subsequently, the testing phase evaluates the project's real-life compatibility through techniques like code correctness, error detection, performance tests, and usability evaluation. Finally, the maintenance phase addresses error elimination and improvements throughout the software's lifecycle [1].

Testing, which accounts for over 50% of developmental costs, is particularly resource-intensive when conducted manually and thus is a vital concern to the software development life cycle [3]. This method, requiring human judgment for each test run, becomes increasingly impractical as system size and complexity grow, especially within rapid release (continuous integration/ continuous delivery) environments [4].

Traditional testing strategies, involving manual testing and Scripted Automation systems like Jenkins, Maven, and Gradle, often face limitations such as incomplete test coverage as testers write test cases based on guesses about customer interaction with the application, which is time-consuming and often inaccurate [5] and a lack of testing data die to privacy and security concerns [6]. Introduction of new Test automation, enhanced by AI technology, mitigates these issues by reducing manual labor and streamlining repetitive tasks. AI-driven automation learns from past results to predict potential issues [7], employing techniques such as big data analysis for reading data and evaluating it from different sources, machine learning for recognizing patterns and unstructured datasets, and artificial neural networks acting as mathematical routines to classify and arrange data. [8]

AI-supported technologies enhance the effectiveness, speed, and efficiency of software development processes, even outperforming actual engineers in certain domains [1]. Generative AI (GenAI) significantly contributes to ideation, automating repetitive tasks such as code generation, testing, and documentation, thus freeing up valuable time for developers [9]. With the rapid adoption of technologies like ChatGPT, which reached 100 million users within just two months [10], the potential for AI to transform software development is evident. ChatGPT, powered by large language models (LLMs) like GPT-4, leverages vast knowledge bases and advanced natural language processing capabilities to perform complex tasks and enhance human-computer interactions [9]. Trained on extensive corpora [11], these models can identify defects, anomalies, and vulnerabilities more accurately, thereby reducing the time and resources required for manual testing and debugging [12]. Moreover, being tailored for conversational use, the ability to generate responses that are intelligent and human-like, makes it produce even more creative and out of the box results [13].

This research goes into depth on the intricacies of software testing and the application of generative AI within this domain, distinguishing it from other literature reviews. Unlike existing papers that often focus on using AI in software testing or employing generative AI in the broader software lifecycle, this study zeroes in on the unique advantages of generative AI specifically for software testing. It highlights why generative AI stands out as the optimal choice compared to other AI technologies by thoroughly examining its challenges and successes. Furthermore, while many papers limit their scope to particular aspects such as test case generation, this paper provides a comprehensive analysis of generative AI across all domains of testing. This holistic approach not only underscores the multifaceted benefits of generative AI but also identifies and addresses gaps in current research, advocating for further work where existing efforts have fallen short.

The remainder of this paper is organized as follows. Section 2 provides a comprehensive technical overview of software testing, detailing the various methodologies and frameworks used. It also introduces generative AI, explaining its unique features and potential applications within testing domains. Section 3 outlines the research methodology employed in this study, detailing the approach taken to gather and analyze relevant data. Section 4 presents the findings, showcasing the potential of generative AI in software testing and reviewing current work in this area. Section 5 offers a detailed discussion, addressing the research questions and highlighting both the advantages and shortcomings of using generative AI in software testing, along with recommendations for future research. Finally, Section 6 concludes the paper, summarizing the key insights and emphasizing the significance of generative AI.

**2. Background**

This section provides a brief overview of the testing components in a software subsystem, detailing the types and methods of testing. It further introduces Artificial Intelligence and large language models, explaining generative AI and the construction of the most popular models. Additionally, it explores how these AI models are utilized in software testing.

*2.1. Test Components in the Software Testing Life Cycle*

A comprehensive test case encompasses a testing prefix, testing input, testing oracle, and a specification, as depicted in Figure 1. A test case begins with a test prefix, which is a series of meticulously planned activities, configurations, or inputs that prepare the system for the test scenario. This includes initializing variables, setting system preferences, and supplying initial data inputs to establish a controlled environment. A well-written test prefix ensures that each test iteration starts from a consistent and well-defined beginning, which is crucial for assuring the repeatability and dependability of test findings [14].   Next is a test input that is the series of information inputted during the test to determine its execution. A test oracle, on the other hand, is a source of information that determines whether the output of a software system is correct or not [15]. There are two categories of possible states for a system under test produced by these oracles: one for states that satisfy all prerequisites and are regarded as operating normally, and another for extraordinary situations where prerequisites have not been met [14]. For example, in a system when testing its capabilities on an oracle it would be a success state and one would be in a failure state.

A test case specification includes the preconditions outlining the system under test's (SUT) environment and condition prior to execution, the test steps detailing the procedures to follow, the predicted results indicating what should happen, and the actual results documenting the outcomes. This detailed approach ensures a comprehensive assessment of the software's functionality and performance under specified conditions [16]. Testing involves defining and running tests, often synonymous with verification and validation, to inform stakeholders about the software product's quality [17]. Effective testing aligns closely with user requirements, and is done within limited time and resources [2] accounting for about 30-40% of a software organization's work and more than 50% the budget.   while involving various team members, and includes proper reporting and logging. This process involves various team members and includes proper reporting and logging to ensure the successful delivery of a high-quality software service or product [16].

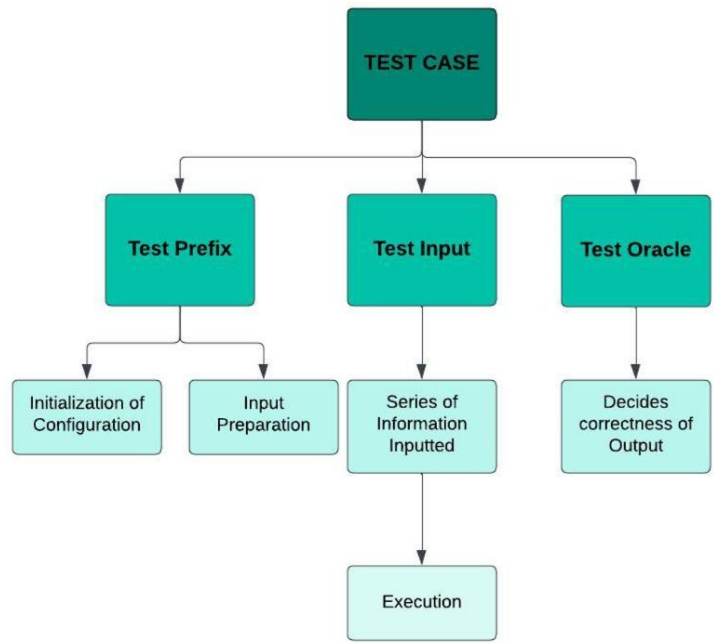

**Figure 1.** Parts of a test case.

*2.2. Testing Strategies and Techniques*

An effective testing system has coverage criteria that determine how much of the code is covered by testing. Key metrics including statement, branch, condition, and path coverage. High coverage indicates that test cases cover a significant portion of the code, which improves fault detection rates, retesting consistency, and overall test execution efficiency. Comprehensive coverage is essential for ensuring that the software functions correctly in various scenarios. There are some key factors to test performance like fault detection rate that helps in assessing the effectiveness of a test case, Retesting rate that helps in testing the consistency of the test rate and execution time that is important to maintain a balance between speed and quantity [17].

The Software Testing Life Cycle (STLC) encompasses phases such as requirements analysis, test planning, test development, test execution, evaluating exit criteria, and test closure [19, 20]. AI approaches have significantly impacted these phases, enhancing efficiency and accuracy [12]. Some currently used approaches used by modern organizations like Agile and Extreme Programming methodologies increase the focus on testing by prioritizing time and fast delivery, necessitating thorough acceptance testing based on user requirements [17].

A software System Under Test (SUT) comprises various testing strategies acting like different parts of a system in different conditions [17]. Starting with the most basic, unit testing verifies individual components of software to ensure they work as intended [1]. Integration testing follows, testing the interaction between integrated components to identify interface defects by using multiple unit tests together [1]. Lastly system testing evaluates the system's compliance with specified requirements, ensuring the entire system functions correctly [1]. Additional tests include User Acceptance Testing (UAT), which end users conduct to ensure the system meets real-world requirements, performance testing, which assesses system performance under various loads, security testing, which ensures data integrity and protection against external threats, and regression testing, which ensures that new changes do not introduce new issues by retesting existing functionality [1]. As highlighted in Figure 2, the different stages of a Software Testing lifecycle are responsible for different stages of Test Techniques. This closely links to the V model Diagram of Software Development but is specific to testing.

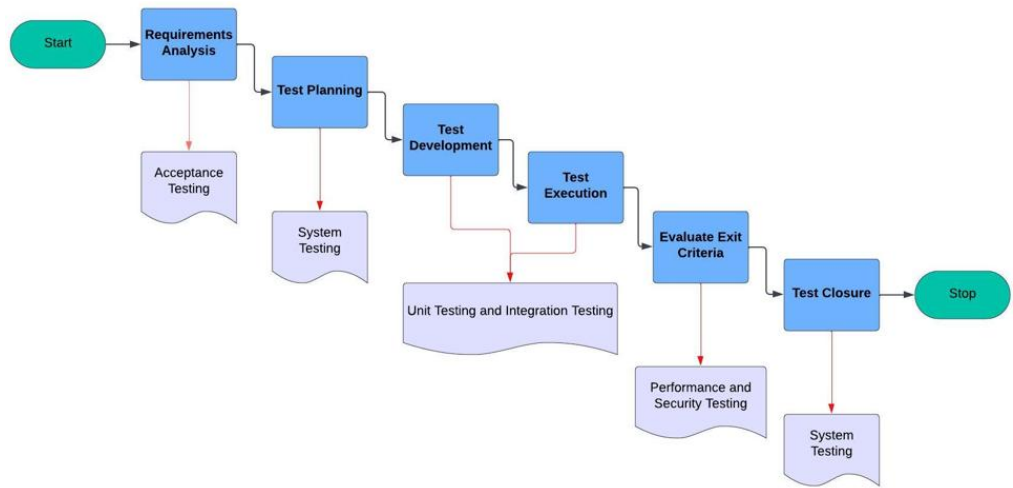

**Figure 2.** Software testing life cycle and testing phases.

A software System uses different Testing Practices to implement these Strategies mentioned above and create a fully functioning Testing System [17]. Modern testing practices such as Test-Driven Development (TDD), Acceptance Test Driven Development (ATDD), and Story Test Driven Development (STDD) emphasize testing as a core property and prioritize unit tests [17].

The choice of testing strategy depends on the availability of information and the specific requirements of the software being tested [17]. There are various testing techniques including White Box Testing that analyzes the program's code logic with full access to the source code, allowing testers to understand and test the internal workings [21]. Black Box Testing that tests the application based on its output without knowledge of the internal workings, focusing on the software's external behavior [21]. Gray-Box Testing which combines both approaches, studying the modules for creating test cases and testing exposed interfaces, providing a balanced approach to testing. Gray-testing involves studying the modules for the purpose of creating test cases (white-box), but the real testing is done in the exposed interfaces (black-box) [21].

*2.3. AI in Software Testing*

The term Artificial Intelligence (AI) was first coined by John McCarthy at a Dartmouth Conference and refers to systems where machines simulate intelligence [16]. AI involves creating intelligent machines, especially computer programs, capable of human-like behavior and integrating new information into existing data structures to make inferences. This capability enables AI to automate complex tasks and make decisions based on vast amounts of data, enhancing efficiency and accuracy in various domains, including software testing [8].

Various forms of AI systems are used in software testing. Firstly, Predictive AI uses historical data to predict future outcomes by collecting information from ongoing processes and improving the model and application under test. This is implemented in software testing to identify patterns and trends that indicate failures and risks, enhancing the testing system and ensuring user satisfaction and product quality [22]. Primarily predictive AI (deep learning models) are used in security testing and maintenance as a software testing tool for devices [22]. The main challenge that comes along with the use of Predictive AI agents is the availability of the dataset. To use an agent to predict information for the future, it needs to have a backup of an old data set that is used to help train the model for the future. It also needs the development of heavy forecasting regression models to predict scenarios. Still, the results are not the most accurate [23]. Next type of AI system is called Generative AI as it is used to make new processes, refine and maintain the ongoing ones, and generate required objects from natural language [24]. Generative AI can be used in almost all domains of SDLC like requirements gathering, Software Design, Coding, and Testing. For software testing, it is used to generate test cases, and inputs/outputs pairs to validate them with the test cases [2]. Incorrect prompts/ not asking for the right information is one of the biggest challenges that generative AI faces. Basically, when the prompt is misleading the generated test or input/output pair may be irrelevant [24]. Lastly, the Assistive AI model which uses a machine learning tool and is used to further assist and automate the decisions and actions towards a task is fairly popular in Engineering. This is done by letting the AI make decisions instead of human interference [25]. It is a great tool for maintenance as it can monitor and easily work on changes. It works well with constantly changing user requirements and is great in highlighting abnormalities and risks making it a good tool to test for failures [26]. One huge risk would be security and loss of control. As the AI system has access to every little piece of information to work on and make decisions, it ends up having a higher degree of control than the programmers themselves [25].

Thus, Generative AI showcasing a specialized broader spectrum makes it the basis of this research. Unlike the other AI systems that primarily process or analyze existing data, generative models work in the territory of creation. The distinctive feature lies in using their capabilities for generating new testing scenarios and expanding the product similarly.

*2.4. Large Language Models*

The motivation behind using generative AI is its ability to synthesize or generate responses to queries. Unlike traditional search engines that require users to sift through numerous results, generative AI Transformers provide the best matching framed responses to requests [9].

Generative AI models, or Large Language Models (LLMs), are based on transformer architecture. The fundamental element of the transformer model is the self-attention mechanism, which enables the model to focus on several portions of the input sequence simultaneously. The transformer architecture includes an encoder and a decoder, each with multiple layers of feedforward neural networks and self-attention. This enables the model to consider the entire input sequence at once when generating predictions. The input sequence is processed by the encoder, while the output sequence is produced by the decoder. Among other things, the transformer's self-attention method enables the model to focus on pertinent segments of the input sequence, making it easier to capture long-range dependencies and enhance translation quality [27].

The two most popular transformer-based generative AI models are ChatGPT and BERT. ChatGPT (Chat Generative Pre-Trained Model Transformer) predicts the next word in a sequence based on the previous context, allowing for the creation of fluent and contextually relevant text. It is pre-trained on a large corpus of text, including web pages and books [27]. The GPT model was released by OpenAI in 2018, and subsequently, versions (GPT2, GPT-3, GPT-3.5, and GPT-4) were released. Many observers saw a notable step shift in generating performance with GPT-3 and 3.5, which sparked a lot of interest in LLMs in general and GPT (and ChatGPT) in particular [11]. ChatGPT functions as a chatting or assisting agent, requiring prompts as input data, which it processes alongside its database [1]. Tasks such as code generation, test automation, bug analysis, and performance improvements have been performed by ChatGPT [1]. GPT's Codex which is a Transformer specifically designed to work for generating code is used in software engineering tasks [27]. On the other hand, BERT (Bidirectional Encoder Representation Transformer) randomly masks words in the input text and trains the model to predict the masked words based on their context [27]. It learns to determine whether two sentences in each document are consecutive or not [27]. BERT comes pre-trained on a large corpus of text, such as Wikipedia and Book Corpus, and uses unsupervised learning and large-scale transformer architectures to capture general language representations [27]. Specialized generative systems like CuBERT and CodeBERT, designed for code generation, are used for software-specific tasks [27].

## 3. Methodology

This section outlines our review methodology and related information, including our research questions and objectives, search strategy and sources, inclusion and exclusion protocols, and the data collection and extraction process.

### 3.1. Research Question

The research questions addressed in the study are as follows:
- RQ1. What are the various types of AI in software testing?
- RQ2. How effective is generative AI in assisting software engineers with software testing?
- RQ3. What is the potential of incorporating generative AI into software testing processes?
- RQ4. What are the advantages and drawbacks associated with the use of generative AI in software testing?

### 3.2. Objectives

This comprehensive analysis aims to explore and document the various ways in which generative AI is currently being used, assessing its effectiveness in enhancing software testing practices compared to traditional methods. Additionally, this research seeks to investigate the diverse applications and effectiveness of generative AI in software testing, along with its potential, limitations, and drawbacks. In addition to these aims, the study intends to identify current research gaps to direct future research efforts. The possible benefits of using generative AI in software testing will also be discussed, including better test case creation, quicker problem detection, increased test coverage, and overall cost savings. Simultaneously, it will examine the disadvantages and restrictions related to the application of generative AI, such as issues with data privacy, the difficulty of implementation, the requirement for large datasets, and potential biases in AI models. Ultimately, this research aims to provide a thorough understanding of the role of generative AI in software testing, offering insights into its current state and challenges, and outlining areas for future exploration and improvement.

### 3.3. Search Strategy

To ensure a comprehensive review, we focused on the most relevant and high-quality papers. Our search strategy included accessing specific databases like Google Scholar and online libraries to find publication resources such as SpringerLink, ScienceDirect, ResearchGate, arXiv, and IEEE Xplore. The search terms and strings used to identify pertinent papers were as follows:
- ("" OR "Challenges of" OR "Benefits of") AND ("Generative AI in " OR "Large Language Model in" "Natural language processing in") AND ("Software Testing" OR "Software Quality" OR "Continuous Integration")

- "Use of AI in Software Development Life Cycle"
- "Types of AI Techniques in Software Testing"
- "Conventional Software Testing Techniques"

*3.4. Inclusion & Exclusion Criteria*

We established our inclusion and exclusion criteria (see Table 1) based on the scope of our research. Our sources included websites, peer-reviewed journals, blogs, and research papers, as well as expert opinions. We excluded articles not written in English and those not targeting the use of AI in software testing.

Table 1. Inclusion and exclusion criteria of the papers.

| Inclusion Criteria | Exclusion Criteria |
|---|---|
| English | Languages other than English |
| Related to Generative AI in Software Testing | Related to other uses of Generative AI |
| Research Papers, Peer-reviewed Articles, Blogs, and Websites. | Social-Media and Other Similar Platform Articles. |
| Personal Opinion from an Expert in the field. | Not focused on Software Testing |
| From 2000-2024 | Duplicate studies across different databases |

*3.5. Data Collection and Extraction Process*

The data extraction process adhered to a structured three-tier workflow including Searching, Refining, and Addition stages. Initially, databases such as Google Scholar, Science Direct, Arxiv, and IEEE Xplore were queried, and the retrieved documents were filtered through the Inclusion and Exclusion Criteria. The Refining stage was pivotal, involving careful examination of titles, abstracts, and full document contents, leading to multiple eliminations. Data was categorized into two classes: topic-specific, consisting of research exclusively related to Generative AI in Software Testing, and topic-related, including General AI concepts potentially applicable to Software Testing. The final stage involved the addition of relevant papers sourced from references, which underwent further refinement, as shown in Figure 3.

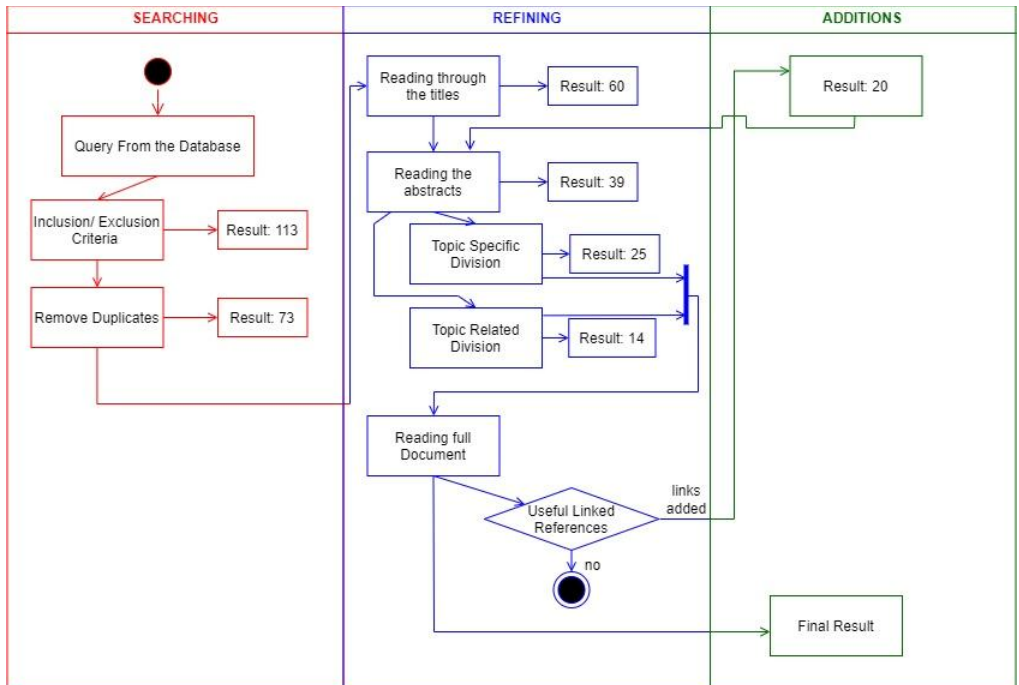

**Figure 3.** Overview of data collection process.

A high-level overview of the data collection and extraction process is depicted in Figure 4 to illustrate the approach taken to ensure the selection of relevant and high-quality papers. Initially, papers were eliminated based on their titles, ensuring that only those directly related to the research topic were considered. Subsequently, abstracts were carefully reviewed to further filter out irrelevant studies. The remaining papers were then classified into specific categories, allowing for a more organized and targeted extraction of information. This categorized information was meticulously mapped to provide a comprehensive understanding of the current landscape of generative AI in software testing.

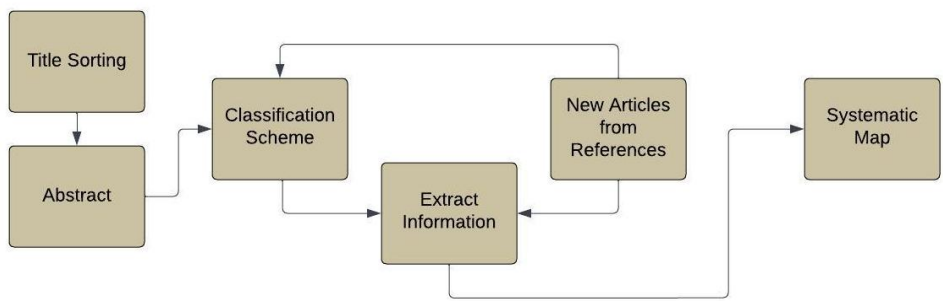

**Figure 4.** Overview of the data collection and extraction process.

The 59 included papers are represented in the line graph in Figure 5, which illustrates a notable increase in research publications following the advent of OpenAI in 2019. This trend highlights the burgeoning interest in generative AI for software testing, particularly in recent years. The graph effectively demonstrates how the majority of the research activity has surged post-2019, indicating a growing recognition of the potential and relevance of generative AI in this domain. This surge in interest not only underscores the contemporary significance of the topic but also suggests a promising trajectory for future advancements and higher levels of success in integrating generative AI into software testing practices.

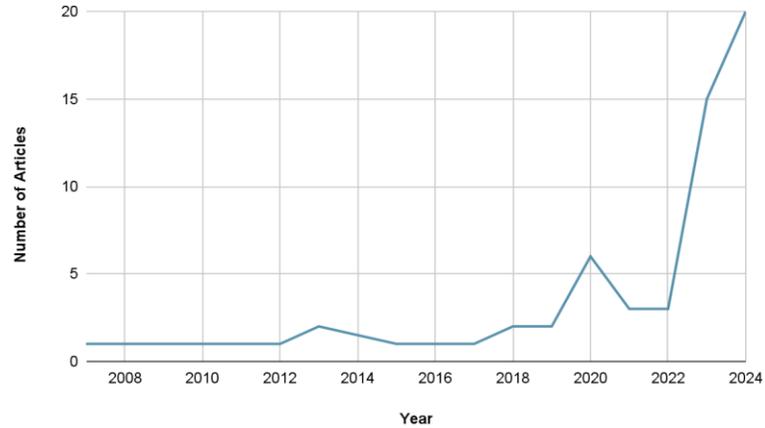

**Figure 5.** Number of publications in each year.

## 4. Findings

Following the methodology outlined above, a total of 59 papers were identified as relevant to the scope of this research paper. These papers were selected based on their alignment with the research focus, ensuring a comprehensive coverage of Generative AI in Software Testing. The selection process involved meticulous evaluation of each paper's contribution to the field, its relevance to the research objectives, and its potential to enhance the understanding of the topic. This selection process ensures that the papers included in this research paper are not only pertinent but also contribute significantly to the discourse on the subject matter.

*4.1. Potential of Generative AI*

Pre-trained models and Large Language Models (LLMs) excel at language representations and contextual awareness, adapting even better through fine-tuning, transfer learning, and prompt engineering [17]. Many generative AI systems are already integrated into the software development cycle. For example, GitHub Copilot, which uses OpenAI's Codex, provides code completion by integrating directly within Integrated Development Environments (IDEs). Recently, the release of GPT-4-powered Copilot X has further enhanced this capability, offering generative question answering within the context of the code [9]. This potential makes AI-driven testing an optimal choice for modern software development. According to Wang et al., current research highlights that software testing can be effectively performed by fine-tuning LLMs or employing prompt engineering with generative AI [28]. The immense potential of AI, particularly Natural Language Processing (NLP) and generative AI, is currently being harnessed in several domains of software testing. These include test prioritization, predicting manual test case failures, generating test cases from software requirements, automatically documenting unit test cases, and detecting coverage gaps [19]. However, there are certain areas where AI may be less efficient or where traditional techniques still prevail, such as test data generation and black-box testing [19]. This section of the paper will present various methods for optimizing generative AI for software testing, exploring how different technologies can be employed to perform distinct software testing tasks effectively.

4.1.1. Fine tuning Large Language Models (LLMs) for efficient generation

Fine-tuning a Large Language Model (LLM) is a critical process that involves adjusting a pre-trained model to perform specific tasks more efficiently. This process allows the LLM to better understand and generate outputs that align with the desired outcomes for particular applications, such as software testing. Fine-tuning leverages additional data and training techniques to refine the model's capabilities, resulting in improved accuracy, speed, and relevance of generated content.

Extensive work has been done in fine-tuning LLMs to enhance their efficiency. Algarsamy et al., made their model called A3Test, that was first fine-tuned with assertions and focal context and then the tuned

LLM was used to generate different results for testing. The results of fine tuning showed that it was 147% more correct and 97.2% faster than other pre-trained generative transformers [29]. Another work was AthenaTest by Tufano et al. This model was a method of fine tuning an LLM on the basis of developer code rather than assertions or context and was primarily designed for white box and mutation testing. It seeks to provide unit test cases by studying test cases written by developers and real-world focal techniques. They use a two-step training procedure that involves denoising pre-training on a sizable unsupervised Java corpus and supervised fine-tuning of the downstream job of producing unit tests. They frame unit test case generation as a sequence-to-sequence learning challenge [30]. Another great example of fine tuning would be CodeMosa. It is based on Search-based software testing to improve coverage and then use OpenAI codex to generate code. This is constant monitoring of coverage of a System under test and making sure it's in the specified criteria. Codex is used to generate arbitrary code and if SBST tests the generative code is correct and would increase the code coverage, it is included as a part of the test base [31].

4.1.2. The Importance of Prompt Engineering in Enhancing Generative A1

The ability to work with AI effectively and efficiently has become crucial with the widespread adoption of generative AI. Therefore, prompt engineering—a systematic process of crafting prompts or inputs to generative AI models to elicit useful outputs—is essential knowledge for users of generative AI [13]. Prompt engineering involves using fine-tuned prompt words to enhance an LLM model's ability to identify and repair faulty code [17]. This need arises because human languages are inherently ambiguous, which can lead to mistakes or misinterpretations in human-machine interactions [13].

The base LLM model's capacity to generate patches is limited by its code processing capabilities, as it is not trained further. Therefore, prompt engineering is necessary to increase the efficiency of LLMs [17]. Additionally, studies conducted by Fan et al. have shown significant improvements in the generation of software programs, with performance increases ranging from 50% to 80% on models like CodeX, CodeGeeX, and CodeGen across different programming languages such as Python, Java, and JavaScript [11].

Previously, Xie et al. developed a system called ChatUniTest, which operates based on prompt-engineered generation. This study uses a methodology known as focal point generation, where as much context related to the topic is included in the prompt while staying within the token limit. This optimizes the prompt for better generation within the token constraints. The solution also includes features such as syntactic, runtime, and compilation validation with automatic repair. The performance, based solely on prompt-engineered inputs, achieved a code coverage of around 59.6%, compared to other fine-tuned LLMs that achieved around 38-42% [32]. Another example of prompt-engineered usage is MuTAP by Dakhel et al., which is used to verify mutation testing. The primary purpose of the tool is to re-prompt the generated results to enhance their effectiveness. The platform performs better than other test generation tools like Pynguin and standard LLM generations. This methodology can detect about 17% of human-written buggy code, which no other proposed solution has achieved yet [33]. Another survey by Schäfer et al. introduced TestPilot, an adaptive JavaScript test-generating tool that can create unit tests for methods based on OpenAI's GPT-3.5-turbo LLM. TestPilot embeds contextual information about the function being tested into the prompt, such as its signature, documentation comments, usage examples from the project documentation, and the function's source code. The model performs better than many competing models using only sufficient prompting rather than fine-tuning or pre-training the LLM. Additionally, when a generated test fails, TestPilot adaptively constructs a new prompt encapsulating the test and failure messages to assist the model in resolving the problematic test [34] TestPilot also includes features like Document Miner, Prompt Generator, Test Validator, and Prompt Refiner, achieving 70.2% code coverage and 52.8% branch coverage [34]. Further research by Yuan et al., Guilherme et al., and Li et al. has also demonstrated the use of prompt engineering in generating unit tests through various methods and validating and fixing them. These studies highlight the versatility and effectiveness of prompt engineering in leveraging generative AI for software testing [15].

4.1.3. Other methods to improve testing systems with generative AI

To further improve AI's effectiveness in software testing, researchers have explored various innovative methods beyond traditional techniques. These methods include leveraging large language models (LLMs) and prompt engineering to optimize the generation of test cases and improve overall testing efficiency. This section will discuss several notable approaches in this domain, showcasing the potential of these advanced methods in enhancing AI-driven software testing.

Based on a combination of LLM and prompt engineering, Patrick et al. developed a 3-step model. First, they create prompt instances by adding some generated prompts with test case examples. Second, the generated prompt is entered into Codex. Lastly, post-processing is done to remove duplication and unmatched test cases. They find that this strategy compares favorably to feedback-directed test creation in a restricted examination of 18 Java methods [35]. Other studies focus on using documentation for test case generation, leveraging natural language requirement descriptions to generate test cases. This approach is advantageous because it is easy to articulate, specify, verify, and use natural language requirements [36]. One research highlights an idea to preprocess the requirements, tag them with parts-of-speech (POS), and parse them using parsers trained to process general-purpose natural language text. Each parsed tree generated by the parser is converted to a knowledge representation graph, which is then combined to generate a single graph representing the knowledge conveyed by the requirement. Finally, this method uses the graph to generate test cases [36]. Another work by Vikram et al., PBTGPT, uses similar API documentation and ChatGPT to generate property-based tests [37]. For oracle generation via documentation, TORADOCU extracts exception oracles for test cases from Javadoc comments. This approach uses a Javadoc extractor to list exceptions that can occur and convert them into natural language using NLP. The generated oracle shows a coverage of over 75% and mostly above 90%, compared to around 50% usually achieved by developers [38]. TestPilot, developed by Schäfer et al., is another effective tool that automatically creates unit tests for methods based on a project's API documentation [34]. Furthermore, an automated framework for providing feedback to enhance a new test case specified in natural language rather than generating it for the developer has been proposed. This framework focuses on three main areas: recommending improvements for the test case description, identifying potential missing steps, and suggesting similar existing test cases [39].

*4.2. Exploration in Testing Domains*

The following Table 2. present below highlights how different researches that contribute to different parts of Software testing domains and link to generative AI as a whole.

**Table 2.** Testing domains and specific researches.

| Testing Domain | Current Use in Generative AI | Research Projects or Work |
|---|---|---|
| Test Case Generation | ✓ | Rajbhog et al. [42] Siddiq et al. [43] Lahiri et al. [44] |
| Test Oracle Generation | ✓ | Tufano Et al. [30] Goffi et al. [38] |
| Test Data Generation | | |
| Test Case Prioritization | ✓ | Yang et al. [46] |
| Test Maintenance | ✓ | M Dewey [14] Tufano et al. [31] T. Cser [48] |
| Intelligent Execution | ✓ | Smith and Wei [40] Gao et al. [49] S. D. Konreddy [50] Thummalapenta et al. [51] |

|                        |   | Cui et al. [22] |
|------------------------|---|-----------------|
| Improved Bug Detection | ✓ | A. Bakshi [52]  |
|                        |   | Folio [53]      |
| Predicting Failures    | ✓ | Hemmati and Sharifi [4] |
| Result Validation      | ✓ | He et al. [54]  |

4.2.1. Test Case Generation

Machine learning techniques, such as genetic algorithms and reinforcement learning, enable AI systems to analyze codebases, requirements, and historical test data to automatically generate test cases [40]. Natural language processing (NLP) has been found to be effective in generating test cases with an accuracy ranging between 70-90% [41]. Generative AI models can analyze pre-existing software code, specifications, and user requirements, assimilating the system's intricate patterns and logic. This understanding of inputs, outputs, and anticipated behavior allows them to generate test cases that cover both expected and edge scenarios. This automated process reduces manual labor and enhances the thoroughness of the testing process by exploring a wider range of potential inputs and scenarios [5]. One notable example is the Meta Model, which generates functional, non-functional, unit, and system test cases based on input services and screen details derived from design specifications [42]. A study by Rajbhog et al. presents a four-step prompting model for test generation. The first step involves inputting context information, such as the business domain and design specifications. The subsequent steps progressively refine the test cases until they are fully generated by the end of the fourth cycle. This method successfully generated and executed 85 test cases for screening (UI), services, and system testing [42]. Another study compares the effectiveness of three large language models (LLMs)—Codex, CodeGen, and ChatGPT—in generating unit test cases for a project. This comparison involved engineering prompts by embedding them with context, signatures, documentation, and related source code [43]. Additionally, a study by Lahiri et al. introduced an interactive test case generation framework based on the CodeX LLM. This framework completes the function body, natural language description, and function header/signature with method names and parameters based on the user's request. The LLM generates a set of candidate codes and tests, and the system then selects a test and asks the user if it aligns with their intent. The user's response is used to prune, sort, and modify suggestions for existing code and test sets. When the interaction ends, the system generates a user-approved set of tests and a ranking [44].

4.2.2. Test Oracle Generation

Test oracle generation is a critical aspect of software testing that involves determining whether the outcomes of a test are correct. This process can be significantly enhanced using machine learning and natural language processing techniques. Tufano et al. refined their pre-trained model to include assertions and focal methods, which improved the generation and exact match ratio of the oracle generated, achieving an accuracy of up to 62% [30]. Another notable work in oracle generation via documentation is TORADOCU, which extracts exception oracles for test cases from Javadoc comments. This system utilizes a Javadoc extractor to compile a list of potential exceptions, which are then converted into natural language using NLP techniques. The generated oracle consistently shows coverage rates above 75%, often exceeding 90%, compared to the typical 50% coverage achieved by developers [38].

4.2.3. Test Data Generation

Test data generation is essential for evaluating software applications under various conditions. It involves creating datasets that can effectively test the functionalities and performance of a system. It has been observed that, in the realm of test data generation, genetic algorithms are more prominent than generative AI. While natural language processing and machine learning algorithms have shown some capability, they have primarily been limited to producing data for GUI testing [45]. This highlights a significant research gap in the application of generative AI for comprehensive test data generation, underscoring the need for further exploration and development in this area.

#### 4.2.4. Test Case Prioritization

Test case prioritization is crucial for identifying the most effective test cases to execute early, ensuring that critical issues are uncovered promptly. The first study introduces three strategies for test case prioritization: The first is the Risk strategy (Risk), which dynamically selects the riskiest test case in each iteration. The second is the Diversity strategy (Div), which prioritizes test cases by maximizing distances from previously inspected test cases. The third strategy is a hybrid prioritization strategy (DivRisk), which combines the two previous strategies [46]. Overall, six sets of cases were tested, and it was found that Risk and DivRisk each won in three. Given that the margin of DivRisk's wins was fairly small, the paper concludes that the Risk strategy is more effective for test case prioritization [46]. Another significant achievement in oracle generation and test prioritization is TeCo by Nie et al. This model uses six different code semantics as inputs and performs reranking via test execution. While primarily designed to help developers complete test code more quickly, its robust execution context has proven useful for oracle generation and test case prioritization tasks as well [47].

#### 4.2.5. Test Maintenance (Regression Testing)

Regression testing is essential for maintaining the quality of software by ensuring that changes do not introduce new issues. This process involves retesting the affected areas and confirming that the program functions as intended. Generative AI can integrate various domains into a single, rigorous system, capable of interpreting visuals, underlying code, and human-readable content. This multidimensional approach ensures that written tests are consistently up-to-date, acting as a guardian of test cases. It optimizes the system to ensure that coverage is tested each time there is a change [48]. Regression testing focuses on identifying modifications in the software being tested. By systematically retesting the impacted areas, it aims to minimize the risk of introducing problems when making changes to the software. This criterion enhances the quality of the test suite, ensuring that the product under test meets its requirements [14]. A notable tool in this domain is CodeMosa, which is built on Software-Based Software Testing (SBST). CodeMosa specializes in situations where code coverage stalls. In such cases, it generates hints to find test cases targeting under-covered functions. This tool effectively covers both the generation and maintenance domains of software testing [31].

#### 4.2.6. Intelligent Execution

Test selection is a critical aspect of various testing methodologies across different systems [3]. Generative AI significantly enhances test case refinement through intelligent execution, allowing the model to handle and select the most appropriate test cases and suites according to the codebase. This adaptability ensures that the most relevant and effective tests are executed, even as the codebase evolves [48].

A study focused on test selection and reduction introduced the Multi-Objective Genetic Algorithm (MOGA) for reducing the number of test cases for web applications while achieving maximum coverage with reduced cost, time, and space [49]. The tool initially generates a requirement document, which is then processed using natural language processing to analyze each case. It generates conjunctive phrases with "if" and "then" keywords, which are entered into the test case table. Using a system called Doc2Vec, it classifies test cases by converting natural language into vector-formatted numbers and classifying them based on an algorithm. The tool runs the classified steps to execute tests, comparing actual and predicted results to determine the status as Pass or Fail [50]. Another tool, ATA, takes input from a manual test case and outputs a script mechanically interpretable by a driver that enables the execution of test cases written in natural language. It views natural language test steps as segments describing actions on targets, resolving ambiguities through backtracking until a successful sequence of calls is generated. This technique automates 82% of test case generation [51]. Feedback from test execution also helps refine tests and allows models to learn about their efficiency, further enhancing the testing process [40].

#### 4.2.7. Improved Bug Detection

Generative AI excels at identifying complex software bugs that often elude human testers due to their intricate interconnections, dependencies, and non-linear behaviors. By analyzing vast amounts of software data, including code, logs, and execution traces, generative AI models can uncover hidden patterns and anomalies, indicating potential issues that might otherwise go unnoticed. Recent research shows that bug reports generated by software testers using generative AI tools exhibit over a 40% reduction in inaccuracies compared to those produced through conventional techniques [52, 53]. Bug detection using a tool called muTap. MuTap has three major components, Initial prompt with zero-shot learning, secondly Generation of test cases by LLM then prompt augmentation by mutants [22].

4.2.8. Predicting Failures

Manual test case failure prediction using Natural Language Processing (NLP) significantly enhances the accuracy of traditional history-based predictions. Given that test cases in manual testing are written in natural language, the application of generative AI in this domain is highly beneficial [4]. The proposed model employs a straightforward NLP technique known as Part of Speech (POS) Tagging, which extracts keywords from test cases and weights them using the Term Frequency Inverse Document Frequency (TF-IDF) metric. This approach has been found to improve the efficiency of failure prediction by 24% compared to traditional techniques [4].

4.2.9. Result Validation

Despite advancements in automated test case generation, the determination of a program's passing status largely relies on manual efforts. Langdon et al. proposed a novel approach to address this challenge by leveraging search-based learning from open-source test suites to generate partially correct test oracles. Their method combines mutation testing, n-version computing, and machine learning techniques to accelerate automated output checking and input generation. This approach represents a promising direction for improving the efficiency and reliability of result validation in software testing [54].

## 5. Discussion

The integration of generative artificial intelligence (AI) in software testing has revolutionized traditional testing methodologies, offering unprecedented capabilities in test case generation, oracle generation, test data generation, test selection, and result validation. This discussion explores the implications of these advancements in the context of software testing, highlighting the potential benefits, challenges, and future directions of generative AI in enhancing the quality and efficiency of software testing processes.

*5.1. Why use Generative AI in Software Testing*

Generative artificial intelligence (AI) is increasingly valuable in software testing due to its ability to enhance various aspects of the testing process. One key benefit is the maximization of test case generation through techniques like identifying critical paths, edge cases, and input combinations [40]. This comprehensive approach ensures thorough testing of software functionalities, improving overall test coverage and effectiveness [55]. AI-driven test management further enhances test coverage by leveraging pattern recognition and probabilistic analysis [40]. By generating tests, optimizing planning and execution, and detecting bugs and errors earlier in the developmental stage, AI improves test quality and reduces the likelihood of issues in the final product [12].

Moreover, generative AI contributes to cost savings and availability by streamlining testing processes and reducing the need for manual intervention. It also enhances speed and efficiency through the use of chatbots, which automate test case execution, data collection, and report generation [12]. This automation not only expedites response times but also improves the efficacy of test reporting and execution. Chatbots facilitate testing automation and continuous integration, enabling greater test coverage across various devices, platforms, and configurations [12].

Generative AI offers a unique advantage over traditional techniques by producing tests that are less likely to be duplicates from the training data. This ensures that tests are either unique or similar, enhancing the diversity and effectiveness of the test suite [12]. Additionally, AI software testing improves accuracy and reduces development time compared to traditional methods. Meeting software development deadlines becomes less challenging, especially with the growing demand for rapid software delivery. Another significant benefit of generative AI is its ability to simplify collaboration in geographically dispersed teams [7]. AI systems can handle repetitive, labor-intensive tasks, allowing software testers to focus on tackling complex issues. This collaborative approach ensures that testing remains thorough and efficient, even across distributed teams [7].

In the realm of Software Testing, the use of large language models (LLMs) like GPT 3.5 and GPT-4, along with chain of thought prompting, has shown significant improvements in generating outputs for complex IoT testing scenarios [56]. LLMs can be trained on specific data sets to detect vulnerabilities in IoT devices, enhancing security testing and ensuring the robustness of IoT applications [56].

*5.2. Some limitations with Generative AI in Software Testing*

Even with the increasing demand for AI, it comes with many limitations and challenges. Most importantly the primary concern is the ethical implications of granting autonomous control to AI systems, as it raises questions of accountability due to the actions performed by artificial agents [40]. The ethical dimension involves understanding how decisions made by AI systems impact human lives and who is responsible when these systems fail or cause harm. The unpredictability of AI behavior further complicates accountability, as it can be difficult to pinpoint the origin of errors in highly complex systems. Additionally, issues such as algorithmic bias and data privacy arise, as AI requires extensive training data, potentially violating privacy and introducing biases into the system [40]. Bias in AI can stem from skewed training data or inherent biases in the algorithms themselves, leading to unfair treatment of certain groups. Privacy concerns are heightened when AI systems handle sensitive data, as unauthorized access or data breaches could have severe consequences. Moreover, the vast amount of data required for training sophisticated AI models often includes personal information, raising significant concerns about data protection and consent. Large language models (LLMs) like GPT may amplify biases and errors present in the training data, leading to discriminatory behavior in the models' outputs [55]. Such biases can perpetuate stereotypes and propagate misinformation, thereby undermining the credibility and reliability of AI systems. Furthermore, ensuring the diversity and representativeness of training data is an ongoing challenge that requires continuous monitoring and updating. The development and deployment of generative AI models can be computationally demanding, requiring significant computational power and specialized hardware [57]. This demand not only increases costs but also raises environmental concerns due to the substantial energy consumption associated with training large models. The need for specialized hardware such as GPUs or TPUs also limits accessibility, as smaller organizations may not have the resources to invest in such technology [57].

Transparency is another major concern, as many AI models operate as black boxes, making their decision-making processes opaque and difficult for humans to interpret [57]. This lack of transparency hinders the ability to understand, trust, and effectively manage AI systems. The complexity of these models makes it challenging for users to decipher how inputs are transformed into outputs, which is critical for validating and improving AI systems.

Privacy and ownership of generated content are also significant issues, as content generated by AI models may be based on someone else's work, raising questions about the reliability and ownership of the content [55]. Intellectual property concerns emerge when AI-generated content mimics existing works too closely, leading to potential legal disputes over copyright infringement. This issue is particularly pronounced in creative industries where originality is highly valued. Integrating AI into testing poses additional challenges, requiring advanced technical knowledge and collaboration with stakeholders [58]. Effective integration necessitates a deep understanding of both the AI models and the systems they are testing, as well as seamless coordination among developers, testers, and other stakeholders. This complexity can slow down the adoption of AI-driven testing methodologies and increase the risk of miscommunication and implementation errors. AI's reliance on large amounts of data can result in incomplete test coverage, as it may struggle to detect problems beyond its training set [12]. Ensuring accurate testing requires realistic scenarios and diverse user interactions, which can be challenging to capture and incorporate into AI models

[12]. The limited scope of training data means that AI systems might miss rare or unforeseen issues that could critically impact system performance. Another concern is the uncertainty in AI's outputs, responses, or actions, as AI-based models rely on statistical algorithms that can lead to inconsistent results [49]. This unpredictability can undermine user confidence and pose significant risks, especially in critical applications where reliability and consistency are paramount.

The phenomenon of "hallucination" in generative AI can produce illogical or inconsistent outputs, leading to misinformation in software testing [13]. However, this hallucination can also be an asset, as it can reveal unexpected software behavior, benefiting testing efforts [59]. This characteristic benefits all levels of LLM use for testing, as 'hallucination' can happen at any level of content generation when using LLMs. Unintended outputs can highlight hidden flaws and edge cases that might not be evident through conventional testing methods.

Automation complexity is a significant limitation but also an opportunity for AI-driven test automation. While complex training and optimization algorithms are required, the potential benefits of increased speed, accuracy, and coverage outweigh these initial challenges. Successful test automation plans are essential due to the increasing complexity of software testing processes [7]. The initial investment in developing robust AI-driven testing frameworks can lead to long-term efficiencies and improvements in software quality, making it a worthwhile endeavor despite the upfront challenges.

*5.3. Potential for Future Work*

Researchers currently believe that AI cannot fully replace a tester but serves as a force multiplier with the potential to do so in the future [20]. Even after generating test cases, testers need to review and monitor tests to ensure sufficient coverage and add new cases as needed [39]. Open problems in AI testing, such as prompt engineering, augmenting test cases, and rerunning them, still exist [11]. There were several shortcomings in the generation and prioritization domain of testing. One significant issue is that not all test cases generated by LLMs may be executable. A survey found that nearly a quarter of the tests generated by ChatGPT were only executable, a number that increased to a third with suitable prompt engineering [11].

The future potential of generative AI in software testing is immense. One promising area is the development and use of local LLMs, which operate directly on users' machines. This approach addresses significant privacy concerns, as data does not need to be transferred to external servers. Running AI models locally ensures that sensitive information remains within the user's control, reducing the risk of data breaches and unauthorized access. Moreover, local models can be fine-tuned to the specific needs and environments of individual users or organizations, enhancing their relevance and effectiveness.

However, there is a notable lack of research on generating testing data relevant to specific test cases. Current methodologies often rely on large, generalized datasets that may not capture the unique nuances of every application. Advancements in this area could significantly improve the precision and applicability of AI-generated test cases. Custom data generation tailored to particular software contexts would ensure more comprehensive test coverage and better detection of edge cases and rare scenarios.

There is potential for AI to analyze high-risk areas of the code base, prioritize tests accordingly, and manage resource allocation [12]. This capability could transform how testing resources are deployed, focusing efforts on the most critical parts of the code and improving overall software reliability. Programming AI algorithms for simulated testing can significantly improve application code validation by accurately depicting typical scenarios for software testers and recognizing and reproducing all possible scenarios [7]. This would lead to more robust and resilient software systems, as AI can continuously adapt and learn from new data and user interactions.

There is believed to be untapped potential in AI predictive analytics, which can enhance software solutions by identifying potential test cases, increasing reliability, and exceeding customer expectations [7]. Predictive analytics could anticipate future issues based on historical data, allowing preemptive action and reducing the incidence of bugs in production. AI will play a crucial role in testing emerging technologies such as cloud computing, IoT, and big data. The combination of AI and new technologies to create testing data for specific products will drive innovation in AI software testing [7].

Generative AI can facilitate the creation of realistic and varied testing environments that mirror actual usage patterns, enabling more accurate and thorough testing. This capability is particularly valuable for complex systems where manual test case generation is impractical or impossible. As AI continues to evolve, its integration with testing frameworks will become more seamless, providing testers with powerful tools to enhance their workflows and outcomes [7].

The potential applications of AI extend far beyond software testing, encompassing areas such as project planning, implementation, design, and problem analysis [8]. Many experts believe that the current development of AI tools by developers is just the beginning, and there is much more innovation to come [8]. For instance, tools like Socrates utilize automated testing techniques and a specific middleware to bridge gaps in the testing process. These tools act as automated guides, assisting testers rather than replacing them, thus enhancing the testing process and ensuring more reliable software [59]. As AI continues to evolve and integrate into various facets of software development, its role will become increasingly indispensable, driving efficiency, accuracy, and innovation across the industry.

*5.4. Threats to Validity*

This section highlights the potential threats to validity of our research from Internal and External factors.

5.4.1 Internal Threats

The internal threats to the validity of this research are primarily based on the researcher's selection bias and database query results of the reference papers. Other influencing factors include, but are not limited to, researcher bias, the categorization process, and the inclusion and exclusion criteria. To mitigate these issues, several steps were taken. A high-level refining process was used, beginning with title-based elimination, followed by sorting papers as abstracts were read. This was followed by fully reading and extracting information from the selected papers. Additionally, a process of adding papers through thorough reference checking was implemented, leading to a comprehensive search across the topic. The majority of the search was based on factual data and evidence present in the studies, with opinions being avoided. However, some threats may arise due to the general nature of the study, as it is not extremely specific to a single topic but is a wide area of research.

5.4.2. External Threats

The external threats to the validity of this study are primarily related to the coverage of the referenced papers. Measures were taken to address this issue, as previously mentioned, by thoroughly examining the references of the selected papers and including additional relevant studies. Given the significant interest in the use of AI across various fields, an effort was made to include and cover every relevant study related to the topic. However, there is no complete guarantee that the process and elimination criteria did not miss any relevant studies. Despite these efforts, the comprehensive nature of the topic makes it challenging to ensure that all relevant papers were included.

## 6. Conclusion and Future Work

This research successfully addresses all the research questions. RQ1 is explored in the Background section, which discusses Predictive, Generative, and Assistive AI Models. RQ2 is covered throughout the paper, with particular emphasis on the effectiveness of Generative AI in the Findings and Discussions sections, demonstrating the extensive work done in fine-tuning and prompt engineering, as well as the areas requiring further exploration. The potential of Generative AI in software testing i.e., RQ3 is highlighted in the Findings, showcasing its capabilities in tasks such as test case generation, oracle generation, and prioritization and more. Lastly, RQ4 is thoroughly examined in the Discussions, where the challenges and advantages of using Generative AI in testing are explained in detail.

In conclusion, integrating Generative AI into software testing provides numerous advantages, including enhanced test coverage, faster and more efficient testing processes, and cost savings. However, several challenges and limitations, such as ethical concerns, algorithmic bias, and data privacy issues, must be addressed. These challenges underscore the need for careful consideration and responsible implementation of AI in testing practices.

Despite these obstacles, the potential of Generative AI in software testing is vast. As technology advances, AI will increasingly enhance testing processes, especially in areas like IoT testing and test case generation. This paper has highlighted the extensive implementation of AI in certain areas, such as test generation and validation, while also noting the limited support Generative AI offers in domains like testing data. There is a need for more research, particularly in using Generative AI for testing data generation. Future efforts should focus on overcoming current limitations and exploring new opportunities to leverage AI for further improvement in software testing practices on local systems rather than losing control. Overall, adopting Generative AI has the potential to transform software testing, making it more efficient, effective, and reliable.